\documentclass[twocolumn,showpacs,preprintnumbers,amsmath,amssymb]{revtex4}

\usepackage{graphicx}
\usepackage{dcolumn}
\usepackage{bm}

\usepackage{amssymb}
\usepackage{amsmath}

\begin{document}


\title{ Non-adiabatic-like accelerated expansion of the late universe in entropic cosmology }

\author{Nobuyoshi {\sc Komatsu}$^{1}$}  \altaffiliation{E-mail: komatsu@t.kanazawa-u.ac.jp} 
\author{Shigeo     {\sc Kimura}$^{2}$}

\affiliation{$^{1}$Department of Mechanical Systems Engineering, Kanazawa University, 
                          Kakuma-machi, Kanazawa, Ishikawa 920-1192, Japan \\
                $^{2}$The Institute of Nature and Environmental Technology, Kanazawa University, 
                          Kakuma-machi, Kanazawa, Ishikawa 920-1192, Japan}%
\date{\today}

\begin{abstract}

In `entropic cosmology', instead of a cosmological constant $\Lambda$, an extra driving term is added to the Friedmann equation and the acceleration equation, 
taking into account the entropy and the temperature on the horizon of the universe.
By means of the modified Friedmann and acceleration equations, 
we examine a non-adiabatic-like accelerated expansion of the universe in entropic cosmology.
In this study, we consider a homogeneous, isotropic, and spatially flat universe, focusing on the single-fluid (single-component) dominated universe at late-times.
To examine the properties of the late universe, we solve the modified Friedmann and acceleration equations, neglecting high-order corrections for the early universe. 
We derive the continuity (conservation) equation from the first law of thermodynamics, 
assuming non-adiabatic expansion caused by the entropy and temperature on the horizon. 
Using the continuity equation, we formulate the generalized Friedmann and acceleration equations, and propose a simple model.  
Through the luminosity distance, it is demonstrated that the simple model agrees well with both the observed accelerated expansion of the universe and a fine-tuned standard $\Lambda$CDM (lambda cold dark matter) model. 
However, we find that the increase of the entropy for the simple model is likely uniform, while the increase of the entropy for the standard $\Lambda$CDM model tends to be gradually slow especially after the present time.  
In other words, the simple model predicts that the present time is not a special time, unlike for the prediction of the standard $\Lambda$CDM model.

\end{abstract}

\maketitle

\section{Introduction}
Numerous cosmological observations have implied a new paradigm for the cosmic expansion history, i.e., an accelerated expansion of the universe \cite{Tyson1988,PERL1998a,PERL1998b,Riess1998,Riess2004,Riess2007,SN1,Tegmark1,Spergel1,Wood2007,Kowalski1,Hicken1,EKomatsu1,WMAP2011}.
To explain the accelerated expansion, an additional energy component called `dark energy' is usually added to 
both the Friedmann equation and the Friedmann--Lema\^{i}tre acceleration equation, where general relativity is assumed to be correct. 
In particular, $\Lambda$CDM models, which assume cold dark matter (CDM) and a cosmological constant $\Lambda$, have been suggested as an elegant description of accelerated expansion \cite{Fukugita01,Carroll01,Ryden1,Hartle1,Sato1,Weinberg1,Roy1}. 
(For other models, see Refs.\ \cite{Weinberg1,Roy1,Miao1,Bamba1} and references therein.)

Recently, Easson, Frampton, and Smoot \cite{Easson1,Easson2} have proposed that an extra driving term  should be added to the Friedmann--Lema\^{i}tre acceleration equation.
The additional entropic-force term can explain the accelerated expansion of the late universe \cite{Easson1} and the inflation of the early universe \cite{Easson2}, without introducing new fields \cite{Koivisto1}. 
In the entopic-force scenario, called `entropic cosmology', 
the additional driving term is derived from the usually neglected surface terms on the horizon of the universe in the gravitational action, 
assuming that the horizon has an entropy and a temperature \cite{Easson1}.
(In fact, the entropy and temperature are related to the Bekenstein entropy \cite{Bekenstein1} and the Hawking temperature \cite{Hawking1} of black holes on an event horizon.)
Since then, many researchers have extensively examined entropic cosmology from various viewpoints \cite{Koivisto1,Cai1,Cai2,Qiu1,Casadio1,Danielsson1,Myung1,Costa1,Basilakos1}. 
(The possibility that the entropic force on the horizon can explain the accelerating universe \cite{Easson1} should be distinguished from the idea that gravity itself is an entropic force \cite{Padma1,Verlinde1}.)  

In entropic cosmology, since the entropy on the horizon is assumed, the entropy can increase during the evolution of the universe.
Therefore, it is possible to consider that the evolution of the universe is a kind of non-adiabatic process, unlike in standard cosmology, in which an adiabatic (isentropic) expansion is assumed.
Nevertheless, such a non-adiabatic-like expansion of the universe has not yet been extensively investigated in entropic cosmology and has been considered in only a few studies \cite{Cai1,Cai2, Qiu1,Casadio1,Danielsson1}. 
Therefore, it is important to examine the non-adiabatic-like (hereafter non-adiabatic) process, to acquire a deeper understanding of entropic cosmology, especially from a thermodynamics viewpoint.
Also, after the discovery of black hole thermodynamics \cite{Bekenstein1,Hawking1}, 
the entropy of the universe was examined by many researchers \cite{Davies1,Davies2,Davies3,Frautsch1,Barrow1,Barrow2,Sugimoto1}.
In particular, since the late 1990s, the entropy of the universe has been extensively discussed in a universe undergoing accelerated expansion \cite{Easther1,Barrow3,Davies11,Davis01,Davis00,Jaeckel01,Gong00,Gong01,Setare1,Cline1}.
However, evolution of the entropy has not been studied in entropic cosmology, although entropy plays an important role.

In this context, we examine a non-adiabatic expansion of the universe and discuss the evolution of the entropy in entropic cosmology. 
For this purpose, we derive the continuity (conservation) equation from the first law of thermodynamics, 
taking into account the non-adiabatic process caused by the entropy and the temperature on the horizon.
If the modified Friedmann and Friedmann--Lema\^{i}tre acceleration equations are used, the continuity equation can be derived from the two equations without using the first law of thermodynamics. 
This is because two of the three equations are independent \cite{Ryden1}.
However, in this study, we derive the continuity equation from the first law of thermodynamics, since the first law is the fundamental conservation law.
Using the obtained continuity equation, we formulate the generalized Friedmann and Friedmann--Lema\^{i}tre acceleration equations. 
In addition, we propose a simple model based on the formulation. 
It should be noted that we do not discuss entropic inflation \cite{Easson2,Cai2} in the early universe, 
since we focus on the late universe \cite{Easson1,Gao1} to examine the fundamental properties of the universe in entropic cosmology.

The present paper is organized as follows. 
In Sec. II, we give a brief review of the two modified Friedmann equations in entropic cosmology.
In this section, we examine the properties of the single-fluid dominated universe.
In Sec. III, we derive the modified continuity equation from the first law of thermodynamics, 
assuming a non-adiabatic expansion of the universe.
We also discuss generalized formulations of entropic cosmology and propose a simple model. 
In Sec. IV, we compare the simple model with the observed supernova data and several $\Lambda$CDM models.
Finally, in Sec. V, we present our conclusions.

\section{Modified Friedmann equations}
Koivisto {\it et al.} \cite{Koivisto1} have summarized two modified Friedmann equations
to examine the entropic cosmology proposed by Easson \textit{et al.} \cite{Easson1,Easson2}.
In this study, we employ the two modified Friedmann equations.
We do not derive the two modified Friedmann equations in the present paper, 
since the theoretical derivation has been described in Refs.\ \cite{Easson1,Easson2}.
In Sec. \ref{Modification of the Friedmann equations}, we first give a brief review of the two modified Friedmann equations.
In Secs. \ref{Solutions for single fluid} and \ref{Properties}, 
we examine the solutions and the properties of the single-fluid dominated universe in entropic cosmology.

\subsection{Two modified Friedmann equations with additional driving terms for entropic cosmology} 
\label{Modification of the Friedmann equations}

We consider a homogeneous, isotropic, and spatially flat universe, 
and examine the scale factor $a(t)$ at time $t$ in the Friedmann--Lema\^{i}tre--Robertson--Walker metric. 
In entropic cosmology, the two modified Friedmann equations \cite{Koivisto1} are summarized as 
\begin{equation}
 \left(  \frac{ \dot{a}(t) }{ a(t) } \right)^2  = H(t)^2     =  \frac{ 8\pi G }{ 3 } \rho(t)  + \alpha_{1} H(t)^2  + \alpha_{2} \dot{H}(t) ,  
\label{eq:mFRW01(H4=0)}
\end{equation}
and
\begin{align}
  \frac{ \ddot{a}(t) }{ a(t) } &=  \dot{H}(t) + H(t)^{2}                                                                                     \notag \\
                                        &=  -  \frac{ 4\pi G }{ 3 } ( 1 +  3w) \rho(t)   + \beta_{1} H(t)^2  + \beta_{2} \dot{H}(t)   , 
\label{eq:mFRW02(H4=0)}
\end{align}
where the Hubble parameter $H(t)$ is defined by
\begin{equation}
   H(t) \equiv   \frac{ da/dt }{a(t)} =   \frac{ \dot{a}(t) } {a(t)}  .
\label{eq:Hubble}
\end{equation}
$G$ and $\rho(t)$ are the gravitational constant and the mass density of cosmological fluids, respectively.
Note that we neglect high-order terms for quantum corrections, since we focus on the late universe.
In Eq.\ (\ref{eq:mFRW02(H4=0)}), $w$ represents the equation of state parameter for a generic component of matter, 
which is given as  
\begin{equation}
  w = \frac{ p(t) } { \rho(t)  c^2 }    ,
\label{eq:w}
\end{equation}
where $c$ and $p(t)$ are the speed of light and the pressure of cosmological fluids.
For non-relativistic matter (or the matter-dominated universe) and relativistic matter (or the radiation-dominated universe),  $w$ is $0$ and $1/3$, respectively.
In Eqs.\ (\ref{eq:mFRW01(H4=0)}) and (\ref{eq:mFRW02(H4=0)}),  
the four coefficients $\alpha_1$, $\alpha_2$, $\beta_1$, and $\beta_2$ are dimensionless constants \cite{Koivisto1}.
The $H^2$- and $\dot{H}$-terms with the dimensionless constants correspond to the additional driving terms,  
which take into account the entropy and temperature on the horizon of the universe due to the information holographically stored there  \cite{Easson1}.
In this study, Eqs.\ (\ref{eq:mFRW01(H4=0)}) and (\ref{eq:mFRW02(H4=0)}) are called the modified Friedmann equation and the modified (Friedmann--Lema\^{i}tre) acceleration equation, respectively.
[Equation (\ref{eq:mFRW01(H4=0)}) corresponds to energy conservation.]

Easson {\it et al.} \cite{Easson1} have derived the modified acceleration equation, i.e., Eq.\ (\ref{eq:mFRW02(H4=0)}).  
In their paper, the dimensionless constants were expected to be bounded by $\frac{3}{2 \pi} \lesssim \beta_{1} \leqslant 1$ and $0 \leqslant \beta_{2} \lesssim \frac{3}{4 \pi} $. 
Typical values for a better fitting were $\beta_{1} = \frac{3}{2 \pi}$ and $\beta_{2} = \frac{3}{4 \pi}$ \cite{Easson1}. 
It was argued that the extrinsic curvature at the surface was likely to result in 
something like $\alpha_1 = \beta_1 = \frac{3}{2 \pi}$ and $\alpha_2 = \beta_2 = \frac{3}{4 \pi}$ \cite{Easson2,Koivisto1}.
Of course, it is difficult to derive four unknown dimensionless constants from first principles.

We now examine the two modified Friedmann equations.
Coupling [$(1+3w) \times $ Eq.\ (\ref{eq:mFRW01(H4=0)})] with [$2 \times $ Eq.\ (\ref{eq:mFRW02(H4=0)})] and rearranging, 
we obtain 
\begin{equation}
 \dot{H}   =  -  \frac{  3(1 + w)  -  \alpha_{1}(1+3w) - 2\beta_{1}  }{   2 - \alpha_{2}(1 + 3w)  - 2\beta_{2}  }  H^2    . 
\end{equation}
The above equation can be rewritten as 
\begin{equation}
 \dot{H}  = \frac{ dH }{ dt }  = -  C_{1} H^2  , 
\label{eq:dHC1}
\end{equation}
where
\begin{equation}
  C_{1} = \frac{  3(1 + w)  -  \alpha_{1}(1+3w) - 2\beta_{1}  }{   2 - \alpha_{2}(1 + 3w)  - 2\beta_{2}  } .  
\label{eq:C1}
\end{equation}
From Eq.\ (\ref{eq:dHC1}),   $(dH/da) a$ is calculated as
\begin{align}
\left ( \frac{dH}{da} \right )  a     &=       \left ( \frac{dH}{dt} \right )   \frac{dt}{da} a  
                                                    =    ( - C_{1} H^2)   \frac{a}{\dot{a}}  = -  C_{1} H^2  \frac{1}{H}                  \notag \\
                                                 &=  - C_{1}  H   . 
\label{eq:dHda_C1H}
\end{align}
Accordingly, Eq.\ (\ref{eq:dHda_C1H}) is arranged as
\begin{equation}
  \frac{ d H }{ d N } =  - C_{1} H   ,
\label{eq:dHdN(H4=0)}
\end{equation}
where $N$ is defined by 
\begin{equation}
   N  \equiv \ln a   \quad \textrm{and therefore} \quad  dN   = \frac{da}{a}   .  
\end{equation} 
As discussed in Ref.\ \cite{Koivisto1}, the two modified Friedmann equations can be arranged as a simple equation, Eq.\ (\ref{eq:dHdN(H4=0)}).
In the next subsection, we solve Eq.\ (\ref{eq:dHdN(H4=0)}), assuming a single-fluid dominated universe.

\subsection{Solutions for the modified Friedmann equations in the single-fluid dominated universe}
\label{Solutions for single fluid}
We can solve Eq.\ (\ref{eq:dHdN(H4=0)}) analytically, when $C_{1}$ is constant.
In fact, as shown in Eq.\ (\ref{eq:C1}), $C_{1}$ is constant when $\alpha_{i}$, $\beta_{i}$, and $w$ are constant values.
(Here $\alpha_{i}$ represents $\alpha_{1}$ and $\alpha_{2}$ and $\beta_{i}$ represents $\beta_{1}$ and $\beta_{2}$.) 
Therefore, to solve Eq.\ (\ref{eq:dHdN(H4=0)}), we assume that $\alpha_i$ and $\beta_i$ are constant.
In addition, for a constant $w$, we assume the single-fluid dominated universe.
Concretely speaking, $w$ is $0$ and $1/3$ for the matter- and radiation-dominated universes, respectively.

When $C_1$ is constant, Eq.\ (\ref{eq:dHdN(H4=0)}) can be integrated as
\begin{equation}
 \int \frac{dH}{H} = - \int C_{1} dN   .  
\end{equation}
This solution is given by 
\begin{equation}
  \ln  H  =  - C_{1} N   + D'  =   - C_{1} \ln a + D'  ,  
\end{equation}
and therefore we find  
\begin{equation}
H   =  D a^{- C_{1}}      , 
\label{eq:Solve1}
\end{equation}
where $D'$ and $D$ are integral constants.
Dividing Eq.\ (\ref{eq:Solve1}) by $H_{0} = D a_{0}^{- C_{1}} $, we have   
\begin{equation}
 \frac{ H }{ H_{0} }   =  \left ( \frac{ a } {  a_{0} } \right )^{ -C_{1}}   ,
\label{eq:H/H0}
\end{equation}
where $H_0$ and  $a_0$ are the present values of the Hubble parameter and the scale factor.
We obtain the above simple solution, 
since we assume the single-fluid dominated universe and neglect high-order terms for quantum corrections.

Equation (\ref{eq:H/H0}) indicates that $C_{1}$ is an important parameter for discussing the universe in the present entropic cosmology.
We can determine $C_{1}$ from Eq.\ (\ref{eq:C1}). 
For example, if $\alpha_{i} = \beta_{i} = 0 $, then 
$C_{1}$ for the radiation-dominated universe ($w=1/3$) is $C_{1,r} = 2$,
while $C_{1}$ for the matter-dominated universe ($w=0$) is $C_{1,m} = 3/2$.
When $\alpha_{i} = \beta_{i} = 0 $, the two modified Friedmann equations are the standard Friedmann and acceleration equations, respectively \cite{Weinberg1,Ryden1,Hartle1,Sato1,Roy1}. 
Accordingly, Eq.\ (\ref{eq:H/H0}) for $C_{1,r} = 2$ and $C_{1,m} = 3/2$ agrees with the standard formula for the radiation- and matter-dominated universes, respectively. 
Note that the universe for $C_{1} = C_{1,\Lambda} = 0$ corresponds to the $\Lambda$-dominated universe, as discussed later.

\subsection{Properties of the single-fluid dominated universe in entropic cosmology} 
\label{Properties}
To observe the properties of the single-fluid dominated universe,  
we examine three properties: the scale factor (in Sec. \ref{Scale factor}), 
the luminosity distance  (in Sec. \ref{Luminosity distance}), 
and the entropy on the Hubble horizon (in Sec. \ref{Entropy on the Hubble horizon}). 
In this subsection, we consider $C_{1}$ as a non-negative free parameter.

\subsubsection{Scale factor $a$} 
\label{Scale factor} 

The first important property we examine is the scale factor $a(t)$.
To this end, Eq.\ (\ref{eq:H/H0}) is arranged as 
\begin{equation}
 \tilde{ H }    =   \tilde{a}^{ -C_{1} }  , 
\label{eqA:H/H0}
\end{equation}
where $\tilde{ H }$ and $\tilde{ a }$ are defined by 
\begin{equation}
 \tilde{ H } \equiv \frac{H}{H_0}  \quad  \textrm{and}  \quad  \tilde{ a } \equiv \frac{a}{a_0}   .
\label{eq:def_H_a}
\end{equation}
Multiplying Eq.\ (\ref{eqA:H/H0}) by $\tilde{ a } $, we obtain
\begin{equation}
\tilde{H} \tilde{ a }   =  \frac{1}{ H_{0} } \frac{ d\tilde{a} } {  dt }      =   \tilde{a}^{ - C_{1}  + 1 }     , 
\label{eqA:dadt}
\end{equation}
where $\tilde{H} \tilde{ a }$ is calculated as  
\begin{align}
\tilde{H} \tilde{ a }    &=  \frac{H}{ H_{0} }  \frac{ a }{ a_0}  
                                 = \frac{ \dot{a}/a }{ H_{0} }  \frac{ a }{ a_0}  
                                 = \frac{ \dot{a}  }{ H_{0} a_{0} }                                                                       
                                 = \frac{  a_{0} \frac{d}{dt}  \left ( \frac{a}{a_0} \right )  }{  H_{0} a_{0}  }   \notag \\
                              & = \frac{1}{ H_{0} }  \frac{d \tilde{a} }{dt}     .
\end{align}
Integrating Eq.\ (\ref{eqA:dadt}) and replacing $\tilde{a}$ by $a/a_{0}$, we finally have  
\begin{equation} 
   \frac{a}{a_{0}}  =    \begin{cases}
                            (C_{1} H_{0} t )^{  \frac{1}{C_{1}} }  &   (C_{1} \neq 0) ,        \\
                            \exp[ H_{0}( t - t_{0} ) ]                &   (C_{1}      = 0) ,        \\
\end{cases}
\label{eqA:a-t}
\end{equation}
where $t_{0}$ represents the present time. 
Note that the integral constants are calculated from $\tilde{a} =1$ at $t={t}_{0}$, 
where ${t}_{0}$ is set to be $1/(C_{1}H_{0})$. 
Typical results are given by 
 \begin{equation}
     \frac{a}{a_{0}}   = \begin{cases}
                 \sqrt{ 2 H_{0} t   }                         &   (C_{1} = C_{1,r}               = 2) ,        \\
                 ( \frac{3}{2} H_{0} t   )^{2/3}          &   (C_{1} = C_{1,m}              = 3/2),     \\
                   \exp[ H_{0}( t - t_{0} ) ]                &   (C_{1} = C_{1,\Lambda}   = 0) .      
\end{cases}
\end{equation}
When $\alpha_{i} = \beta_{i} = 0 $, the above three results correspond to the scale factor
for the radiation-, matter-, and $\Lambda$-dominated universes, respectively \cite{Ryden1}.

\begin{figure} [t]  
\begin{minipage}{0.495\textwidth}
\begin{center}
\scalebox{0.32}{\includegraphics{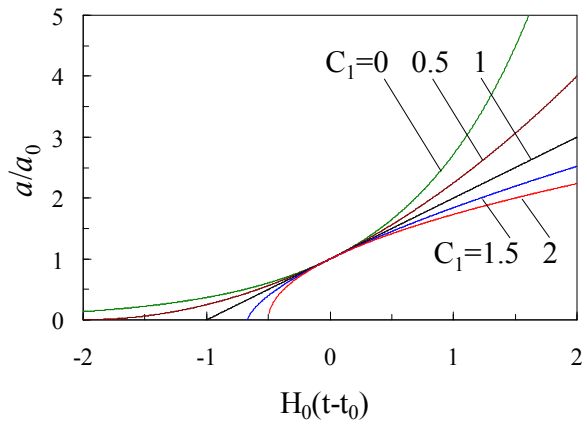}}
\end{center}
\end{minipage}
\caption{ (Color online). Time evolution of normalized scale factor $a/a_{0}$ for single-fluid dominated universe with various $C_{1}$. 
The horizontal axis is normalized as $H_{0}(t-t_{0})$ \cite{Ryden1}. 
The lines for $C_{1} = 2$, $1.5$, $1$, and $0$ are equivalent to those for the radiation-, matter-, empty-, and $\Lambda$-dominated universes, respectively. 
The universe for $C_{1}=0.5$ is different from the above standard single-component universes. }
\label{Fig-a-t_C1}
\end{figure}

Now we consider $C_{1}$ as a non-negative free parameter, to observe the properties of the single-fluid dominated universe.
Time evolutions of the normalized scale factor $a/a_{0}$ with various $C_{1}$ are plotted in Fig.\ \ref{Fig-a-t_C1}. 
The results for $C_{1} = 2$, $1.5$, $1$, and $0$ are consistent with those for the radiation-, matter-, empty-, and $\Lambda$-dominated universes, respectively.
(Entropic-force terms become more dominant as $C_{1}$ decreases.)

As shown in Fig.\ \ref{Fig-a-t_C1}, for $ H_{0}(t-t_{0}) >0$, the increase of the scale factor tends to be faster as $C_{1}$ decreases. 
That is, at late times, the expansion of the universe increases with decreasing $C_{1}$. 
In fact, an accelerated expanding universe is observed when $C_{1} < 1$, e.g., $C_{1}=0.5$ and $0$. 
We can confirm the accelerated expansion from the `deceleration parameter', which is used to discuss the expansion of the universe \cite{Ryden1,Hartle1,Sato1,Weinberg1,Roy1}. 
The deceleration parameter is defined by
\begin{equation}
q_{0} \equiv  - \left ( \frac{\ddot{a} } {a H^{2}} \right )_{t=t_0}  .
\label{eq:q_def}
\end{equation}
Substituting Eq.\ (\ref{eq:dHC1}) into Eq.\ (\ref{eq:mFRW02(H4=0)}), we have 
\begin{equation}
  \frac{ \ddot{a} }{ a }  =  \dot{H} + H^{2}   =  -  C_{1} H^2 + H^{2} = (1- C_{1}) H^2  , 
\end{equation}
and arranging this gives 
\begin{equation}
  \frac{ \ddot{a} }{ a H^2 }  =  1- C_{1}   . 
\label{eq:ddota_a_C1}
\end{equation}
Accordingly, substituting Eq.\ (\ref{eq:ddota_a_C1}) into Eq.\ (\ref{eq:q_def}), the deceleration parameter is given as 
\begin{equation}
q_{0} =  C_{1} -1  . 
\label{eq:q0C1}
\end{equation}
Note that we do not assume the single-fluid dominated universe to calculate $q_{0}$ shown in Eq.\ (\ref{eq:q0C1}).
From Eq.\ (\ref{eq:q0C1}), we find that $q_{0}$ is negative when $C_{1} <1 $.
The negative deceleration corresponds to the acceleration.
That is, when $C_{1} <1 $, the accelerating universe can be mimicked by entropic-force terms. 
Note that Eq.\ (\ref{eq:q0C1}) is different from $q_{0}$ for $\Lambda$CDM models \cite{Weinberg1}, 
which we will discuss in Sec. \ref{Comparison}. 

As mentioned previously, we assumed the single-fluid dominated universe to calculate the scale factor.
However, the universe for $C_{1}=0.5$ is different from the so-called single-component universe, 
such as the radiation-, matter-, empty-, and $\Lambda$-dominated universes appearing in the standard cosmology \cite{Ryden1,Hartle1,Sato1,Weinberg1,Roy1}.
This is because, in entropic cosmology, the entropic-force terms affect the properties of the universe.

\subsubsection{Luminosity distance $d_{L}$} 
\label{Luminosity distance}

The luminosity distance obtained from the observation data has been widely used to study the accelerated expansion of the universe. 
Therefore, we examine the luminosity distance $d_{L}$ of the single-fluid dominated universe in entropic cosmology.
The luminosity distance \cite{Sato1} is given as
\begin{equation}
 d_{L}(z)   = \frac{ c (1+z) }{ H_{0} }  \int_{1}^{1+z}  \frac{dy} {F(y)} ,
\label{eqA:dL-def1}  
\end{equation}
where the integrating variable $y$ and the function $F(y)$ are given by 
\begin{equation}
  y = \frac{a_0} {a} ,   
\end{equation}
and
\begin{equation}
F(y)   = \frac{ H }{ H_{0} } ,
\label{eqA:dL-def2}  
\end{equation}
and $z$ is the redshift defined by 
\begin{equation}
 1 + z \equiv  y = \frac{ a_0 }{ a } .
\label{eqA:z-def}  
\end{equation}

Substituting Eq.\ (\ref{eq:H/H0}) into Eq.\ (\ref{eqA:dL-def2}), and using $y = a_{0}/a$, we obtain $F(y)$ as
\begin{equation}
F(y) = \frac{ H }{ H_{0} }  = \left ( \frac{ a }{ a_{0} } \right )^{ -C_{1}}  = \left ( \frac{ a_{0} }{ a } \right )^{C_{1}} = y^{C_{1}}    .
\label{eqA:FyC1}  
\end{equation}
Substituting Eq.\ (\ref{eqA:FyC1}) into Eq.\ (\ref{eqA:dL-def1}),  and integrating, we have 
\begin{equation}
  \left ( \frac{ H_{0} }{ c } \right )   d_{L}  =  \begin{cases}
         \frac{1+z} {C_{1} -1 } \left [ 1- (1+z)^{ -C_{1} +1 } \right ]   &   (C_{1} \neq 1) ,  \\ 
         (1+z)\ln (1+z)                                                                &   (C_{1} = 1)      , 
\end{cases}
\label{eqA:dLC1}  
\end{equation}
where $C_{1} = 1$ corresponds to the empty-dominated universe \cite{Ryden1}.
Typical results are given by 
 \begin{equation}
  \left ( \frac{ H_{0} }{ c } \right )   d_{L}   = \begin{cases}
                 z                                                                         &   (C_{1} = C_{1,r}             = 2) ,    \\
                2(1+z) \left [ 1- \frac{ 1 }{ \sqrt{1+z} } \right ]         &   (C_{1} = C_{1,m}           = 3/2),  \\
                (1+z)z                                                                   &   (C_{1} = C_{1,\Lambda} = 0)  . 
\end{cases}
\end{equation}

\begin{figure} [t]  
\begin{minipage}{0.495\textwidth}
\begin{center}
\scalebox{0.32}{\includegraphics{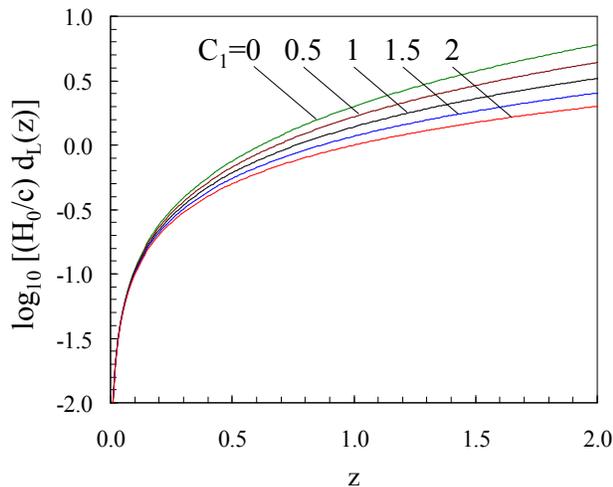}}
\end{center}
\end{minipage}
\caption{ (Color online).  Dependence of luminosity distance $d_L$ on redshift $z$ for single-fluid dominated universe with various $C_{1}$. 
The vertical axis is normalized as $ \log_{10} [ (H_{0} / c ) d_{L} ]$.  }
\label{Fig-dL-z_C1}
\end{figure}

The luminosity distance for various $C_{1}$ is shown in Fig.\ \ref{Fig-dL-z_C1}, where $C_{1}$ is considered as a non-negative free parameter.
As shown in Fig.\ \ref{Fig-dL-z_C1}, the luminosity distance increases with decreasing $C_{1}$, especially in higher $z$ regions.
This indicates that an accelerated expanding universe appears when $C_{1}$ is small. 
In fact, the lines for $C_{1} < 1$, e.g., $C_{1}=0.5$ and $0$, correspond to the accelerating universe, as discussed in Sec. \ref{Scale factor}.
For example, the line for $C_{1} = 0$ is equivalent to the luminosity distance for the $\Lambda$-dominated universe. 
On the other hand, as non-accelerating universes, the lines for $C_{1} = 2$, $1.5$, and $1$ are consistent with those for the radiation-, matter-, and empty-dominated universes, respectively. 
It is clearly shown that $C_{1}$ affects the properties of the single-fluid dominated universe in the present entropic cosmology. 
In Sec. \ref{Comparison}, we will discuss the luminosity distance including the observed supernova data and $\Lambda$CDM models.

\subsubsection{Entropy $S$ on the Hubble horizon} 
\label{Entropy on the Hubble horizon}

In entropic cosmology, we assume that the Hubble horizon has an associated entropy \cite{Easson1}. 
Therefore, as the third property, we examine the entropy on the Hubble horizon.
The Hubble horizon (radius) $r_{H}$ and the entropy $S$ on the Hubble horizon are given as
\begin{equation}
     r_{H} = \frac{c}{H}   , 
\label{eq:rH_00}
\end{equation}
and
\begin{equation}
 S = \frac{ k_{B} c^3 }{  \hbar G }  \frac{A_{H}}{4}   ,
\label{eq:S0_H}
\end{equation}
where $k_{B}$, $\hbar$, and $A_{H}$ are the Boltzmann constant, the reduced Planck constant, and the surface area of the sphere with the Hubble radius $r_{H}$, respectively \cite{Easson1}.
The reduced Planck constant is defined by $\hbar \equiv h/(2 \pi)$, where $h$ is the Planck constant.

Substituting $A_{H}=4 \pi r_{H}^2 $ into Eq.\ (\ref{eq:S0_H}), and using $r_{H}= c/H$ shown in Eq.\ (\ref{eq:rH_00}), we obtain the entropy as
\begin{align}
S
&= \frac{ k_{B} c^3 }{  \hbar G }         \frac{A_{H}}{4}       
   =   \frac{ k_{B} c^3 }{  \hbar G }      \frac{ 4 \pi r_{H}^2 }{4}                                       
   =   \frac{ k_{B} c^3 }{ \hbar  G } \pi   \left ( \frac{c}{H} \right )^2                                     \notag \\
&=  \left ( \frac{ \pi k_{B} c^5 }{ \hbar G } \right )  \frac{1}{H^2}  =  K  \frac{1}{H^2}   ,   
\label{eq:SH}      
\end{align}
where $K$ is a positive constant given by
\begin{equation}
  K =  \frac{  \pi  k_{B}  c^5 }{ \hbar G }     .
\label{eq:K-def}
\end{equation}
For example, the entropy on the Hubble horizon can be evaluated as \cite{Easson1,Egan1}: 
\begin{equation}
      S  \sim     (2.6 \pm 0.3) \times 10^{122}k_{B}  .
\end{equation}
The entropy on the Hubble horizon is far larger than the total of the other entropies of the matter within the horizon \cite{Egan1}.

Multiplying Eq.\ (\ref{eq:SH}) by $H_{0}^2$ and substituting Eq.\ (\ref{eq:H/H0}) into this, we have 
\begin{equation}
 H_{0}^{2} S  = K \frac{ H_{0}^{2} }{H^2}  = K \left ( \frac{ a }{a_{0}} \right )^{2 C_{1}}  , 
\label{eq:H0SH} 
\end{equation}
where the single-fluid dominated universe is assumed since Eq.\ (\ref{eq:H/H0}) is employed.
Substituting Eq.\ (\ref{eqA:a-t}) into Eq.\ (\ref{eq:H0SH}) and rearranging, 
we obtain the entropy $S$ on the Hubble horizon: 
 \begin{equation}
    \left ( \frac{H_{0}^2}{K} \right )  S   = \begin{cases}
                                                                          ( C_{1} H_{0} t )^{2}    &   (C_{1} \neq 0)       ,        \\
                                                                                       1                &   (C_{1}      = 0)       .
\end{cases}
\label{eq:SH-t}
\end{equation}

\begin{figure} [t]  
\begin{minipage}{0.495\textwidth}
\begin{center}
\scalebox{0.32}{\includegraphics{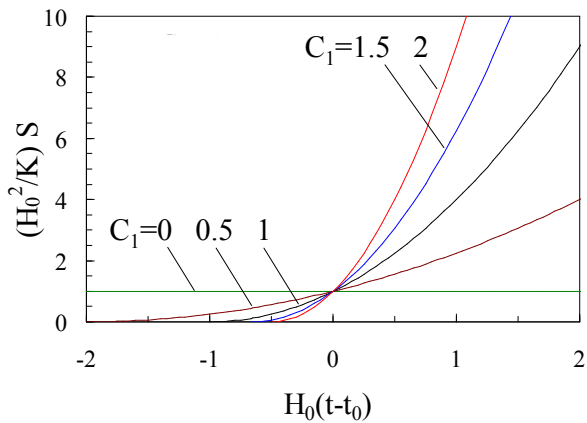}}
\end{center}
\end{minipage}
\caption{ (Color online). Time evolution of entropy $S$ on Hubble horizon for single-fluid dominated universe with various $C_{1}$. 
The vertical and horizontal axes are normalized as $ (H_{0}^2 /K ) S $ and $H_{0} (t -t_{0})$. 
The lines for $C_{1} = 2$, $1.5$, $1$, and $0$ are equivalent to those for the radiation-, matter-, empty-, and $\Lambda$-dominated universes, respectively. 
The universe for $C_{1}=0.5$ is different from the above standard single-component universes.   }
\label{Fig-S-t_C1}
\end{figure}

To observe the entropy, we consider $C_{1}$ as a non-negative free parameter.
The time evolution of the entropy $S$ on the Hubble horizon for various $C_{1}$ is plotted in Fig.\ \ref{Fig-S-t_C1}.
The entropy $S$ for $C_{1}=0$ does not depend on time.
We can confirm this from both Fig.\ \ref{Fig-S-t_C1} and Eq.\ (\ref{eq:SH-t}).
This is because $S$ depends on $H$ as shown in Eq.\ (\ref{eq:SH}), and the Hubble parameter $H$ is constant when $C_{1}=0$. 
(When $C_{1}=0$, Eq.\ (\ref{eq:dHC1}) indicates a constant $H$ because $dH/dt =0$.) 
In contrast, the entropy $S$ increases with time for $C_{1} \neq 0$.  
The increase of entropy is likely consistent with the second law of thermodynamics.
In Sec. \ref{Comparison}, we will discuss the entropy $S$ on the Hubble horizon, including $\Lambda$CDM models.
(Strictly speaking, the other entropies should be taken into account, to examine the generalized second law of thermodynamics as studied in Refs \cite{Davies11,Davis01,Davis00}.  
In the present paper, we do not discuss the generalized second law.)

\begin{figure} [t]  
\begin{minipage}{0.495\textwidth}
\begin{center}
\scalebox{0.32}{\includegraphics{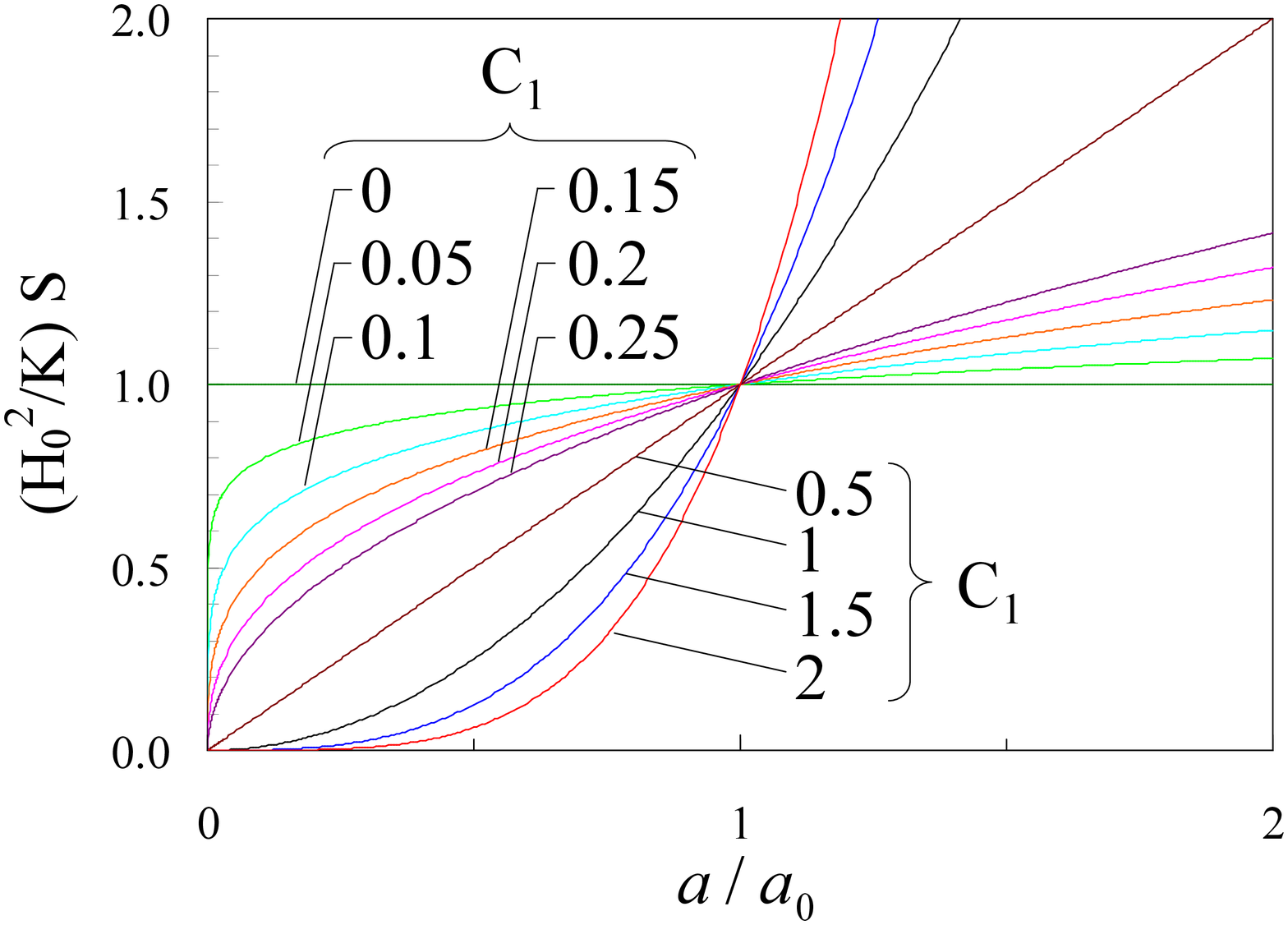}}
\end{center}
\end{minipage}
\caption{ (Color online). Dependence of entropy $S$ on normalized scale factor $a/a_{0}$ for single-fluid dominated universe with various $C_{1}$. 
The vertical axis is normalized as $  (H_{0}^2 /K ) S $.       }
\label{Fig-S-a_C1}
\end{figure}

We now examine the influence of $C_{1}$ on the entropy.
To this end, we observe the dependence of the entropy $S$ on the normalized scale factor $a/a_{0}$, which is given by Eq.\ (\ref{eq:H0SH}).
In Fig.\ \ref{Fig-S-a_C1}, $C_{1}$ is varied from $0$ to $0.25$ with steps of $0.05$, while $C_{1}$ is varied from $0.5$ to $2$ with steps of $0.5$.
As shown in Fig.\ \ref{Fig-S-a_C1}, the entropy rapidly increases with decreasing $C_{1}$, especially for small $a/a_{0}$, e.g., $a/a_{0} < 0.1$, corresponding to early times. 
(Note that we focus on the late universe in this study.)
In contrast, for larger $a/a_{0}$ corresponding to late times, e.g., $a/a_{0} > 1$, the entropy increases slowly as $C_{1}$ decreases.  
In fact, we have observed that the expansion of the universe further accelerates as $C_{1}$ decreases, 
as shown in Eq.\ (\ref{eq:q0C1}) and in Figs.\ \ref{Fig-a-t_C1} and \ref{Fig-dL-z_C1}. 
Therefore, at late times, the entropy increases slowly as $C_{1}$ decreases (or as the entropic-force terms are further dominant), 
while the expansion of the universe accelerates.
We can expect the above relationship between the entropy and the expansion, because $S=K/H^{2}=K/(\dot{a}/a)^{2} $, which is obtained from Eqs.\ (\ref{eq:SH}) and (\ref{eq:Hubble}).  

In this section, we have employed the two modified Friedmann equations \cite{Koivisto1} to study the universe in entropic cosmology. 
We have solved the equations and examined the properties of the single-fluid dominated universe. 
Through the parameter $C_{1}$ related to entropic-force terms, we can summarize the properties of the universe systematically. 
The universe with a specific $C_{1}$ (e.g., $C_{1} = 2$, $1.5$, $1$, and $0$) is consistent with the single-component universe appearing in the standard cosmology.

\section{Formulations of non-adiabatic expansion of the universe}

In the previous section, we examined the properties of the universe described by the  modified Friedmann and acceleration equations.
In this section, we consider a non-adiabatic-like expansion process caused by an entropy and a temperature on the Hubble horizon.
In Sec. \ref{Modified continuity equation}, 
we derive the continuity equation from the first law of thermodynamics,
assuming non-adiabatic expansion of the universe.
In Sec. \ref{Consistency of modified equations}, using the continuity equation, 
we formulate the generalized Friedmann and acceleration equations, and propose a simple model.

It should be noted that several researchers have discussed similar modified continuity equations for entropic cosmology.
For example, Cai {\it et al.} derived the improved continuity equation from the first law of thermodynamics using double holographic screens \cite{Cai1,Cai2}, 
while Qiu {\it et al.} \cite{Qiu1} and Casadio \textit{et al.} \cite{Casadio1} derived the modified continuity equation from the modified Friedmann and acceleration equations.
Danielsson \cite{Danielsson1} examined the sourced acceleration equation using extra source terms, and discussed the modified continuity equation.

\subsection{Modified continuity equation from the first law of thermodynamics for non-adiabatic expansion}
\label{Modified continuity equation}
In this subsection, we derive the modified continuity equation from the first law of thermodynamics, 
assuming non-adiabatic expansion of the universe caused by the entropy and temperature on the Hubble horizon. 
To this end, we first review the continuity equation, according to the textbook by Ryden \cite{Ryden1}.

From the first law of thermodynamics, the heat flow $dQ$ across a region is given by
\begin{equation}
 dQ = dE + pdV         ,
\label{eq:FirstT}
\end{equation}
where $dE$ and $dV$ are changes in the internal energy $E$ and volume $V$ of the region, respectively.
This equation can be rewritten as
\begin{equation}
 dQ = \frac{dQ}{dt} dt = \frac{dE + pdV}{dt} dt  = (\dot{E} + p \dot{V} ) dt    .
\label{eq:FirstT2}
\end{equation}
Let us consider a sphere of co-moving radius $\hat{r}_{s}$ expanding along with the universal expansion 
so that its proper radius $r_{s}(t)$ is given by 
\begin{equation}
 r_{s}(t) = a(t) \hat{r}_{s}   .
\end{equation}
The volume $V(t)$ of the sphere is 
\begin{equation}
 V(t) =  \frac{4 \pi}{3} r_{s}(t)^3      =  \frac{4 \pi}{3} \hat{r}_{s}^3  a(t)^3     ,
\label{eq:V(t)}
\end{equation}
and therefore, the rate of change of the sphere's volume is given as
\begin{equation}
 \dot{V} = \frac{4 \pi}{3} \hat{r}_{s}^3 (3 a^2 \dot{a} ) = V \left ( 3 \frac{\dot{a}}{a} \right )   .
\label{eq:dotV}
\end{equation}
The internal energy $E(t)$ of the sphere is given by 
\begin{equation}
 E(t) = \varepsilon(t)    V(t)      ,
\label{eq:E(t)}
\end{equation}
where the internal energy-density $\varepsilon(t)$ is 
\begin{equation}
  \varepsilon(t) =  \rho(t) c^2  .
\label{eq:varepsilon}
\end{equation}
Differentiating Eq.\ (\ref{eq:E(t)}) with respect to $t$, 
and substituting Eq.\ (\ref{eq:dotV}) into this equation,  
the rate of change of the sphere's internal energy is given as 
\begin{equation}
 \dot{E} =  \dot{\varepsilon} V  + \varepsilon \dot{V} =  \left ( \dot{\varepsilon} + 3 \frac{\dot{a}}{a} \varepsilon \right ) V        .
\label{eq:dotE}
\end{equation}
Substituting Eqs.\ (\ref{eq:dotV}) and (\ref{eq:dotE}) into $ \dot{E} + p \dot{V} $, and using Eq.\ (\ref{eq:varepsilon}), we calculate $ \dot{E} + p \dot{V} $ as  
\begin{align}
 \dot{E} + p \dot{V} 
            &=      \left ( \dot{\varepsilon} + 3 \frac{\dot{a}}{a} \varepsilon \right ) V  
                                                  + p V \left ( 3 \frac{\dot{a}}{a} \right )                                                       \notag  \\
             &=  \left ( \dot{\varepsilon} + 3 \frac{\dot{a}}{a} \varepsilon    +  3 \frac{\dot{a}}{a} p  \right )  V  
                = \left [ \dot{\varepsilon} + 3  \frac{\dot{a}}{a} \left ( \varepsilon   +  p \right )  \right ]  V             \notag  \\           
            &  = \left [ \dot{\rho} + 3  \frac{\dot{a}}{a} \left ( \rho   +  \frac{p}{c^2} \right )  \right ] c^2 V  .        
\label{eq:dotEpdotV}                            
\end{align}
Finally, substituting Eqs.\ (\ref{eq:V(t)}) and (\ref{eq:dotEpdotV}) into Eq.\ (\ref{eq:FirstT2}), 
we obtain the first law of thermodynamics in an expanding (or contracting) universe: 
\begin{align}
  dQ  &= dE + pdV    = (\dot{E} + p \dot{V}) dt                                                                                                \notag  \\  
       &= \left [ \dot{\rho} + 3  \frac{\dot{a}}{a} \left ( \rho   +  \frac{p}{c^2} \right )  \right ] c^2 V dt               \notag  \\ 
       &=\left [ \dot{\rho} + 3  \frac{\dot{a}}{a} \left ( \rho   +  \frac{p}{c^2} \right )  \right ] c^2  \left ( \frac{4 \pi}{3} r_{s}^3  \right )  dt    .     
\label{eq:dQ0}
\end{align}
If we assume adiabatic (and isentropic) processes, then $dQ$ is $0$: that is,  
$dQ = T dS           = 0$,
where $S$ and $T$ represent the entropy and the temperature.
In this case, we obtain the continuity equation for the adiabatic (isentropic) process: 
$ \dot{\rho} +  3 (\dot{a}/a) ( \rho   +  p/c^2 ) = 0 $.

However, in this paper, we examine the universe in entropic cosmology.
That is, the horizon is assumed to have an entropy and a temperature, and therefore, the entropy on the horizon can increase during evolution of the universe.
In summary, we assume a non-adiabatic process given by
\begin{equation}
dQ = TdS \neq 0  .
\end{equation}

To calculate $TdS$, we employ the Hubble radius as the preferred screen, 
since the apparent horizon coincides with the Hubble radius in the spatially flat universe \cite{Easson1}.
The Hubble radius $r_{H}$ is given as
\begin{equation}
     r_{H} = \frac{c}{H}     \quad  \textrm{and therefore}    \quad    \dot{r}_{H}  =  - \frac{ H  \dot{H} }{ c^2 } r_{H}^3 .
\label{eq:rA}
\end{equation}
We assume that the Hubble horizon has an associated entropy $S$ and an approximate temperature $T$ \cite{Easson1}.
The entropy shown in Eq.\ (\ref{eq:S0_H}) is written as 
\begin{equation}
 S = \frac{ k_{B} c^3 }{  \hbar G }  \frac{A_{H}}{4}   ,
\label{eq:S0}
\end{equation}
and the temperature is given by 
\begin{equation}
 T = \frac{ \hbar H}{   2 \pi  k_{B}  } \times  \gamma  =  \frac{ \hbar }{   2 \pi  k_{B}  }  \frac{c}{ r_{H} }   \gamma .   
\label{eq:T0}
\end{equation}
We emphasize that the temperature considered here is obtained from multiplying the so-called horizon temperature, $ \hbar H /( 2 \pi k_{B} ) $, by $\gamma$.  
In this study, $\gamma$ is a non-negative free parameter and is of the order of $O(1)$, typically $\gamma \sim \frac{3}{2\pi}$ or $ \frac{1}{2}$. 
In fact, $\gamma$ corresponds to a parameter for the screen temperature discussed in Refs.\ \cite{Cai1,Cai2,Qiu1}.
Cai \textit{et al.} proposed that cosmological observations constrain the undetermined coefficient \cite{Cai1,Cai2}.
(Easson \textit{et al.} suggested a similar modified coefficient for the temperature \cite{Easson1}.)

The temperature on the horizon can be evaluated as 
\begin{equation}
 T = \frac{ \hbar H}{   2 \pi  k_{B}  }  \times O(1)   \sim  10^{-30} [\textrm{K}]   .  
\end{equation}
The temperature is lower than the temperature of our cosmic microwave background (CMB) radiation \cite{Cai1}, $T_{\textrm{CMB}} = 2.73 [\textrm{K}]$.
Accordingly, strictly speaking, the universe considered here is in thermal non-equilibrium states. 
In the present paper, we assume a non-adiabatic expansion in thermal equilibrium states, using a single holographic screen \cite{Easson1,Easson2}. 
(Thermal equilibrium states in entropic cosmology have been previously discussed using double holographic screens \cite{Cai1,Cai2}.)

From Eqs.\ (\ref{eq:S0}) and  (\ref{eq:T0}), we calculate $TdS$ as 
\begin{align}
 T dS  &= T \times  \frac{ k_{B} c^3 }{  4\hbar G }  \frac{ d A_{H} }{dt} dt 
           = T \times \frac{ k_{B} c^3 }{  4\hbar G }   (8 \pi  r_{H} \dot{r}_{H}) dt  \notag \\
         &= \frac{ \hbar }{   2 \pi  k_{B}  }  \frac{c}{ r_{H} }  \gamma
             \times  \frac{ k_{B} c^3 }{  4\hbar G }   8 \pi  r_{H} \dot{r}_{H} dt    
           = \gamma \frac{c^4}{G} \dot{r}_{H} dt   ,
\label{eq:TdS0}
\end{align}
where  $ dA_{H} /dt$ is   $d(4 \pi  r_{H}^2 )/dt =  8 \pi  r_{H} \dot{r}_{H} $.  
The first law of thermodynamics can be written as 
\begin{equation}
  dQ  = dE + pdV      = TdS     .
\label{eq:dQ_dEpdV_TdS}
\end{equation}
Therefore, substituting Eqs.\ (\ref{eq:dQ0}) and (\ref{eq:TdS0}) into Eq.\ (\ref{eq:dQ_dEpdV_TdS}), we have   
\begin{equation}
       \left [ \dot{\rho} + 3  \frac{\dot{a}}{a} \left ( \rho   +  \frac{p}{c^2} \right )  \right ] c^2  \left ( \frac{4 \pi}{3} r_{H}^3  \right )  dt   
        = \gamma \frac{c^4}{G} \dot{r}_{H} dt     , 
\end{equation}
where the proper radius $r_{s}$ shown in Eq.\ (\ref{eq:dQ0}) is replaced by the Hubble radius $r_{H}$.
Arranging the above equation and substituting Eq.\ (\ref{eq:rA}) into the equation, we obtain
\begin{align}
       \dot{\rho} + 3  \frac{\dot{a}}{a} \left ( \rho   +  \frac{p}{c^2} \right )  
        &=  \gamma \frac{3 c^2}{4 \pi G} \frac{ \dot{r}_{H} }{ r_{H}^3 }    
          =  \gamma \frac{3 c^2}{4 \pi G} \frac{  \left  ( - \frac{ H  \dot{H} }{ c^2 } r_{H}^3 \right )   }{ r_{H}^3 }    \notag  \\
        &= - \gamma \left( \frac{3}{4 \pi G} H  \dot{H} \right )   . 
\label{eq:fluid0}
\end{align}
This is the modified continuity equation derived from the first law of thermodynamics, assuming non-adiabatic expansion of the universe.
The right-hand side of Eq.\ (\ref{eq:fluid0}) is related to the non-adiabatic process. 
If $\dot{H}$ is $0$ or if $H$ is constant, Eq.\ (\ref{eq:fluid0}) is the continuity equation for adiabatic (isentropic) processes.  
We will discuss this in the next subsection.
(A similar improved continuity equation for entropic cosmology has been examined in Refs.\ \cite{Cai1,Cai2}.   
Note that we have derived the modified continuity equation from the first law of thermodynamics, neglecting the entropy for high-order corrections.)

As shown in Eq.\ (\ref{eq:fluid0}), the modified continuity equation has the so-called non-zero term on the right-hand side, as if it were a non-adiabatic process. 
Therefore, in the present paper, we call this the non-adiabatic process. 
(As discussed later, the non-zero term can be cancelled in appearance.)
In fact, it has been known that a similar non-zero term is included in the continuity equation for other cosmological models.  
Accordingly, we introduce two typical models in the following.

The first model is `bulk viscous cosmology', in which a bulk viscosity of cosmological fluids is assumed 
\cite{Davies3,Weinberg0,Murphy1,Barrow11,Barrow12,Zimdahl1,Arbab1,Brevik1,Brevik2,Nojiri1,Meng1,Fabris1,Colistete1,Barrow21,Meng2,Avelino1,Hipo1,Avelino2,Piattella1,Meng3,Pujolas1}. 
Because of the bulk viscosity, a similar non-zero term is included in the continuity equation.
(For bulk viscous cosmology, see, e.g., the work of Barrow \cite{Barrow12}.)
Usually, the only bulk viscosity can generate a classical entropy in homogeneous and isotropic cosmologies. 
However, in this study, we assume an entropy on the horizon of the universe, instead of the classical entropy.
Therefore, in Appendix \ref{Appendix_Bulk}, we discuss similarities and differences between bulk viscous cosmology and entropic cosmology.

The second model is `energy exchange cosmology', in which the transfer of energy between two fluids is assumed \cite{Barrow22}; 
e.g., the interaction between matter and radiation \cite{Davidson1,Szy1}, matter creation \cite{Lima11,Lima12}, 
interacting quintessence \cite{Amendola1,Zimdahl01}, the interaction between dark energy and dark matter \cite{Wang01,Wang02}, 
dynamical vacuum energy \cite{Freese1,Overduin1,Borges1,Carneiro1,Carneiro2,Carneiro3,Fritzsch1}, etc..
In energy exchange cosmology, two continuity equations have a similar non-zero term on each right-hand side.
Note that the two non-zero right-hand sides are totally cancelled, since the total energy of the two fluids is conserved.
For example, using a dynamical vacuum term $\Lambda(t)$  \cite{Borges1,Carneiro1,Carneiro2,Carneiro3,Fritzsch1}, 
the continuity equations for matter `$m$' and vacuum energy `$\Lambda$' should be arranged as 
$\dot{\rho}_{m}           + 3 (\dot{a} / a) ( \rho_{m}            + p_{m}/c^2 )            = -\dot{\Lambda}$  and  
$\dot{\rho}_{\Lambda} + 3 (\dot{a} / a) ( \rho_{\Lambda} + p_{\Lambda }/c^2 ) = \dot{\Lambda}$, respectively.
The continuity equation for matter is equivalent to Eq.\ (\ref{eq:fluid0}), 
if $\Lambda(t)$ is given by $\Lambda  = \sigma H^2$ \cite{Overduin1}, where $\sigma$ is a positive constant.
However, in entropic cosmology, we do not assume a second fluid appearing in energy exchange cosmology.
This is because, in entropic cosmology, an effective continuity (conservation) equation can be obtained from an effective description of the equation of state \cite{Lepe1}, without using a second fluid. 
In this sense, the effective description is likely similar to (single-fluid) bulk viscous cosmology rather than energy exchange cosmology, 
since an effective pressure is employed in bulk viscous cosmology, as shown in Appendix \ref{Appendix_Bulk}.
That is, it is possible to obtain an effective continuity (conservation) equation in appearance, if we employ such an effective description.
For example, when the effective pressure is given by $ p^{\prime} = p + \frac{ \gamma c^2   }{4 \pi G} \dot{H} $, 
Eq. (\ref{eq:fluid0}) is arranged as $\dot{\rho} + 3 ( \dot{a} / a ) ( \rho + p^{\prime} / c^2 )=0$.
In Appendix \ref{Appendix_F}, we discuss the effective description for entropic cosmology. 
Of course, the non-zero right-hand side of Eq.\ (\ref{eq:fluid0}) may be interpreted as the interchange of energy between the bulk (the universe) and the boundary (the horizon of the universe) \cite{Lepe1}, 
as if it were energy exchange cosmology. 
Therefore, it is important to examine a relationship between entropic cosmology and energy exchange cosmology in more detail.
We leave this for the future research.

In the above discussion, we consider $\gamma$ shown in Eq.\ (\ref{eq:fluid0}) as a free parameter for the temperature. 
However, parameters for the entropy, such as Tsallis' entropic parameter, may be required for calculating $TdS$.
This is because nonextensive entropy, e.g., Tsallis' entropy \cite{Tsa0} or Renyi's entropy \cite{Ren1}, has been suggested for generalized entropy of self-gravitating systems 
and has been extensively examined from astrophysical viewpoints \cite{Plas1,Abe1,Torres1,Tsa101,Tsa2,Chava21,Taru4,Naka1,Liu1,Koma2,Koma3,Everton1,Tsallis2012}.
Therefore, not only $\gamma$ but another parameter for the entropy may be required for the modified continuity equation.

\subsection{Generalized formulations and the simple model}
\label{Consistency of modified equations}

We have three modified equations, i.e., the modified Friedmann and acceleration equations [Eqs.\ (\ref{eq:mFRW01(H4=0)}) and (\ref{eq:mFRW02(H4=0)})] and the modified continuity equation [Eq.\ (\ref{eq:fluid0})]. 
Two of the three equations are independent \cite{Ryden1}. 
In this subsection, using the modified continuity equation, we formulate the generalized Friedmann and acceleration equations. 
For this purpose, we select the modified Friedmann equation, Eq.\ (\ref{eq:mFRW01(H4=0)}), and the modified continuity equation, Eq.\ (\ref{eq:fluid0}), as independent equations. 
This is because the two equations are related to the conservation law.
(The modified Friedmann equation corresponds to energy conservation.)

As the two independent equations, the generalized Friedmann equation is given by 
\begin{equation}
 \left(  \frac{ \dot{a} }{ a } \right)^2    =  \frac{ 8\pi G }{ 3 } \rho  + f(t)    ,  
\label{eq:mFRW01(f)}
\end{equation}
and the modified continuity equation is written as
\begin{equation}
       \dot{\rho} + 3  \frac{\dot{a}}{a} \left ( \rho   +  \frac{p}{c^2} \right )  
          = - \gamma \left( \frac{3}{4 \pi G} H  \dot{H} \right )   .
\label{eq:fluid0r}
\end{equation}
Here, $f(t)$ in Eq.\ (\ref{eq:mFRW01(f)}) is a general function related to entropic-force terms including high-order corrections.
Danielsson has examined a similar acceleration equation using an extra source term \cite{Danielsson1}.
(An additional term corresponding to $f(t)$ is not included in the Friedmann equation for bulk viscous cosmology. 
See Appendix \ref{Appendix_Bulk}.)

We now derive the generalized acceleration equation from Eqs.\ (\ref{eq:mFRW01(f)}) and (\ref{eq:fluid0r}).  
To this end, multiplying Eq.\ (\ref{eq:mFRW01(f)}) by $a^2$, we have 
\begin{equation}
 \dot{a}^2     =  \frac{ 8\pi G }{ 3 } \rho a^2 + f a^2    . 
\end{equation}
Differentiating this equation with respect to $t$ gives 
\begin{equation}
 2 \dot{a} \ddot{a}     =  \frac{ 8\pi G }{ 3 } ( \dot{\rho} a^2 + 2 \rho a \dot{a} ) + \dot{f} a^2 +2fa \dot{a}    .
\label{eq:2aa}
\end{equation}
Dividing Eq.\ (\ref{eq:2aa}) by $2 a \dot{a}$ gives 
\begin{equation}
\frac{ \ddot{a} }{ a }   =    \frac{ 4\pi G }{ 3 } ( \dot{\rho} \frac{a}{ \dot{a} } + 2 \rho )   + \frac{1}{2} \dot{f} \frac{a}{\dot{a}} + f    .  
\label{eq:dda_a}
\end{equation}
Multiplying Eq.\ (\ref{eq:fluid0r}) by $a/\dot{a}$ ($=1/H$) and arranging, we have  
\begin{align}
       \dot{\rho}\frac{a}{\dot{a}}  &= - 3 \left ( \rho  +  \frac{p}{c^2} \right )   - \gamma \left( \frac{3}{4 \pi G}  H \dot{H} \right ) \frac{1}{H}   \notag \\ 
                                               &=  - 3( 1 + w)\rho   - \gamma \left( \frac{3}{4 \pi G}  \dot{H} \right )  , 
\label{eq:rho_a_da}
\end{align}
where $w$ is  $p / (\rho  c^2)$ as shown in Eq.\ (\ref{eq:w}).
Accordingly, substituting Eq.\ (\ref{eq:rho_a_da}) into Eq.\ (\ref{eq:dda_a}), 
and using $\dot{a}/a =H$, we obtain  
\begin{align}
\frac{ \ddot{a} }{ a }   
   &=    \frac{ 4\pi G }{ 3 } \left [ - 3( 1 + w)\rho   - \gamma \left( \frac{3}{4 \pi G}  \dot{H} \right )   + 2 \rho \right ]   + \frac{1}{2} \dot{f} \frac{a}{\dot{a}} + f   \notag \\
   &=  - \frac{ 4\pi G }{ 3 } ( 1 + 3w)\rho     + \left ( f   + \frac{1}{2} \frac{ \dot{f} }{H}  - \gamma \dot{H}    \right )            .
\label{eq:accel}
\end{align}
Equation (\ref{eq:accel}) is the generalized acceleration equation, which is derived from Eqs.\ (\ref{eq:mFRW01(f)}) and (\ref{eq:fluid0r}).

Before proceeding further, in this paragraph, we discuss a spatially non-flat ($k \neq 0$) universe, where $k$ is a curvature constant.
To this end, we add $ - k c^2 /a^2$ to the right-hand side of the generalized Friedmann equation, Eq.\ (\ref{eq:mFRW01(f)}), as described in Appendix A of Ref.\ \cite{Easson1}.
In the spatially non-flat universe, the apparent horizon, $r_{A} = c/\sqrt{H^2 + (k /a^2) }$, does not coincide with the Hubble horizon, $r_{H}=c/H$, because of $k \neq 0$.
Accordingly, we employ the apparent horizon as the preferred screen rather than the Hubble horizon \cite{Easson1}.
Consequently, $\dot{H}$ on the right-hand sides of Eqs.\ (\ref{eq:fluid0r}) and (\ref{eq:accel}) should be replaced by $\dot{H} - (k/a^2)$. 
In other words, the modified continuity equation and the generalized acceleration equation should include a curvature term.
Of course, it was argued that the extrinsic curvature at the surface was likely to result in something like $\alpha_1 = \beta_1 = \frac{3}{2 \pi}$ and $\alpha_2 = \beta_2 = \frac{3}{4 \pi}$ \cite{Easson2,Koivisto1}.
Note that we consider a spatially flat ($k = 0$) universe in the present study.

Next, we discuss a simple model.  
For this purpose, we consider only $H^2$-terms as entropic-force terms of the generalized Friedmann equation.   
That is, $f(t)$ in Eq.\ (\ref{eq:mFRW01(f)}) is set to be 
\begin{equation}
  f(t) =  B \times \gamma H^2    , 
\label{eq:fH2}
\end{equation}
where $B$ is a constant. 
(We consider only $H^{2}$-terms, since $\dot{H}$-terms of the modified Friedmann equation are $0$.
The details are summarized in Appendix \ref{Appendix_F}.)
Substituting Eq.\ (\ref{eq:fH2}) into Eq.\ (\ref{eq:accel}), we have 
\begin{align}
\frac{ \ddot{a} }{ a }                                                                            
   &=  - \frac{ 4\pi G }{ 3 } ( 1 + 3w)\rho      + B \gamma H^2     + \frac{1}{2} \frac{ d( B \gamma H^2) / dt  }{H}     - \gamma \dot{H}       \notag \\           
   &=  - \frac{ 4\pi G }{ 3 } ( 1 + 3w)\rho      + B \gamma H^2     + (B- 1) \gamma \dot{H}    . 
\label{eq:accel2}
\end{align}
In fact, Easson {\it et al.} first proposed that the entropic-force terms are $H^2$ or $\frac{3}{2 \pi} H^2$, 
i.e., $\dot{H}$-terms are not included in the entropic-force terms \cite{Easson1}.
Accordingly, we determine $B$ so that the $\dot{H}$-term in Eq.\ (\ref{eq:accel2}) is cancelled.
In other words, for the simple model, we select $B$ as
\begin{equation}
     B =1  \quad \textrm{and therefore} \quad   f(t) =   \gamma H^2    .
\end{equation}
In this case, we can obtain the simple self-consistent equations. 
The simple modified Friedmann, acceleration, and continuity equations are summarized as 
\begin{equation}
 \left(  \frac{ \dot{a} }{ a } \right)^2    =  \frac{ 8\pi G }{ 3 } \rho  + \gamma H^2   ,  
\label{eq:mFRW01(f)3}
\end{equation}
\begin{equation}
\frac{ \ddot{a} }{ a }  =  - \frac{ 4\pi G }{ 3 } ( 1 + 3w)\rho            + \gamma H^2  , 
\label{eq:accel3}
\end{equation}
\begin{equation}
       \dot{\rho} + 3  \frac{\dot{a}}{a}  ( 1   +  w )  \rho   
          = - \gamma \left( \frac{3}{4 \pi G} H  \dot{H} \right )     .
\label{eq:fluid0r3}
\end{equation}
The entropic-force term $\gamma H^{2}$ in Eq.\ (\ref{eq:mFRW01(f)3}) is the same as the term in Eq.\ (\ref{eq:accel3}). 
The above two modified Friedmann equations, i.e., Eqs.\ (\ref{eq:mFRW01(f)3}) and (\ref{eq:accel3}), 
correspond to Eqs.\ (\ref{eq:mFRW01(H4=0)}) and (\ref{eq:mFRW02(H4=0)}) for $\alpha_{1}=\beta_{1} = \gamma$ and $\alpha_{2}=\beta_{2} = 0$.
($\alpha_{1} =\beta_{1}$ satisfies the constraint $ \beta_{1} - \alpha_{1} = 0.02 \pm 0.08$, as is concluded in Ref.\ \cite{Koivisto1}.)
Therefore, we can easily calculate the properties of the single-fluid dominated universe, as examined in Sec. \ref{Properties}.

In the present study, we do not assume a cosmological constant $\Lambda$ and dark energy, since additional driving terms can be derived from the entropic-force on the Hubble horizon. 
However, if $\gamma H^2$ is fixed as a constant $\Lambda /3 $, 
the above three equations are equivalent to $\Lambda$CDM models \cite{Weinberg1}.
(The right-hand side of Eq.\ (\ref{eq:fluid0r3}) is $0$ when $\gamma H^2$ is constant, 
because $d(\gamma H^2)/dt = 2 \gamma H \dot{H} = 0$.)
For example, if we consider the matter-dominated universe with $w=0$ for $\alpha_1=\beta_1=\gamma =1$ and $\alpha_2=\beta_2 = 0$, 
then $C_{1}$ is $0$ from Eq.\ (\ref{eq:C1}).
The universe for $C_{1}=0$ corresponds to the $\Lambda$-dominated universe in the standard cosmology  \cite{Ryden1,Hartle1,Sato1,Weinberg1,Roy1}.

In this section, we have derived the modified continuity equation from the first law of thermodynamics, assuming a non-adiabatic expansion of the universe. 
Using the obtained continuity equation, we have formulated the generalized Friedmann and acceleration equations, i.e., Eqs.\ (\ref{eq:mFRW01(f)}) and (\ref{eq:accel}). 
Moreover, as a possible model, we have proposed the simple model given by Eqs.\ (\ref{eq:mFRW01(f)3}), (\ref{eq:accel3}), and (\ref{eq:fluid0r3}).
We will discuss the properties of the simple model in the next section.
Note that Cai {\it et al.} \cite{Cai1,Cai2}, Qiu {\it et al.} \cite{Qiu1}, Casadio \textit{et al.} \cite{Casadio1}, and Danielsson \cite{Danielsson1} have discussed similar modified continuity equations.
The above obtained equations are related to their works.

\section{Evolution of the late universe in the simple model}
\label{Comparison}

In this section, we examine evolution of the late universe using the simple model given by Eqs.\ (\ref{eq:mFRW01(f)3}), (\ref{eq:accel3}), and (\ref{eq:fluid0r3}). 
To this end, we first examine the luminosity distance $d_L$
because, through $d_L$, we can easily compare the simple model not only with $\Lambda$CDM models but also with the observed supernova data.

For the present simple model, we consider the matter-dominated universe since, in entropic cosmology, we do not assume the cosmological constant and dark energy.
(The influence of radiation is extremely small in the late universe, as discussed later.) 
The matter-dominated universe is given by 
\begin{equation}
  w=0 . 
\label{eq:w=0}
\end{equation}
In the simple model, the four parameters shown in Eqs.\ (\ref{eq:mFRW01(H4=0)}) and (\ref{eq:mFRW02(H4=0)}), 
i.e., $\alpha_1$, $\alpha_2$, $\beta_1$, and $\beta_2$, are set to be 
\begin{equation}
 \alpha_1=\beta_1=\gamma       
\end{equation}
and 
\begin{equation}
 \alpha_2=\beta_2 = 0   ,  
\end{equation}
where  $\gamma$ is assumed to be
\begin{equation}
 \gamma =\frac{3}{2\pi}     \quad  \textrm{or}  \quad    \frac{1}{2}     .  
\label{eq:gamma32pi}
\end{equation}
Substituting the above equations into Eq.\ (\ref{eq:C1}), $C_{1}$ is calculated as
\begin{equation}
        C_{1}  =  \begin{cases}
                         \frac{3}{2}( 1 - \frac{3}{2\pi})= 0.7838 \dotsi        &   (\gamma =\frac{3}{2\pi}) ,  \\ 
                         \frac{3}{4} = 0.75                                                &   (\gamma =\frac{1}{2})     . 
\end{cases}
\label{eq:C1_gamma}
\end{equation}
From Eq.\ (\ref{eqA:dLC1}), we can calculate the luminosity distance $d_L$ for the simple model.
Of course, we accept that $\gamma$ should be a free parameter. 
However, we determine $\gamma$ as shown in Eq.\ (\ref{eq:gamma32pi}). 
(The coefficient $3/(2 \pi)$ was anticipated from the surface term order, while the coefficient $1/2$ was expected from the Hawking temperature description, as described in Ref.\ \cite{Easson1}.)
In fact, the properties of the universe for $\gamma=3/(2 \pi)$ are almost the same as the properties for $\gamma=1/2$, 
since the difference of $C_{1}$ between $\gamma=3/(2 \pi)$ and $1/2$ is small, as shown by Eq.\ (\ref{eq:C1_gamma}).  
Therefore, for the present simple model, we will observe the properties of the universe for $\gamma=1/2$, i.e., $C_{1}=3/4$. 

For $\Lambda$CDM models, the luminosity distance of the spatially flat universe is given as 
\begin{align}
 \left ( \frac{H_0}{c} \right )   d_{L} & =  (1+ z) \int_{0}^{z} dz' [ (1+z')^2 (1+ \Omega_{m} z')  \notag  \\
                                                   & \quad    -z'(2+z') \Omega_{\Lambda}]^{-1/2}   , 
\label{eq:dL(CDM)}
\end{align}
where $\Omega_{m} = \frac{\rho_{m}}{\rho_c} = \frac{8\pi G \rho_m}{3H_{0}^2}$ and $\Omega_{\Lambda}= \frac{\Lambda}{3 H_{0}^2}$ \cite{Carroll01}.
$\Omega_{m}$ and $\Omega_{\Lambda}$ represent the density parameters for the matter and the cosmological constant, respectively. 
$\rho_{c}$ represents the critical density, while $\rho_{m}$ is the density for matter which includes baryon and dark matter.
For the flat universe, $\Omega_{total}$ is given as $\Omega_{total} = \Omega_{m} + \Omega_{\Lambda} =1$. 
Here, we neglect the density parameter $\Omega_{r}$ for the radiation, since $\Omega_{r}$ is extremely small, e.g., $10^{-4} \sim 10^{-5} $ \cite{Ryden1}. 
As typical universes, $(\Omega_{m}, \Omega_{\Lambda})$ is set to be $(1, 0)$, $(0.27, 0.73)$, and $(0, 1)$.
Note that the notation is simplified as $\Omega_{m}= 1$, $\Omega_{m}= 0.27$, and $\Omega_{m}=0$. 
The universes for $\Omega_{m}= 1$ and $0$ correspond to the matter- and $\Lambda$-dominated universes, respectively.
That is, the universes for $\Omega_{m}= 1$ and $0$ are equivalent to the universes for $C_{1}=1.5$ and $0$, as discussed in the previous section.
The universe for $(\Omega_{m}, \Omega_{\Lambda}) = (0.27, 0.73)$ is a fine-tuned standard $\Lambda$CDM model, which takes into account the recent WMAP best fit values \cite{WMAP2011}. 
We numerically calculate $d_L$ for the standard $\Lambda$CDM model, 
since we can not analytically solve Eq.\ (\ref{eq:dL(CDM)}), except for special cases \cite{Carroll01}.

\begin{figure} [t] 
\begin{minipage}{0.495\textwidth}
\begin{center}
\scalebox{0.31}{\includegraphics{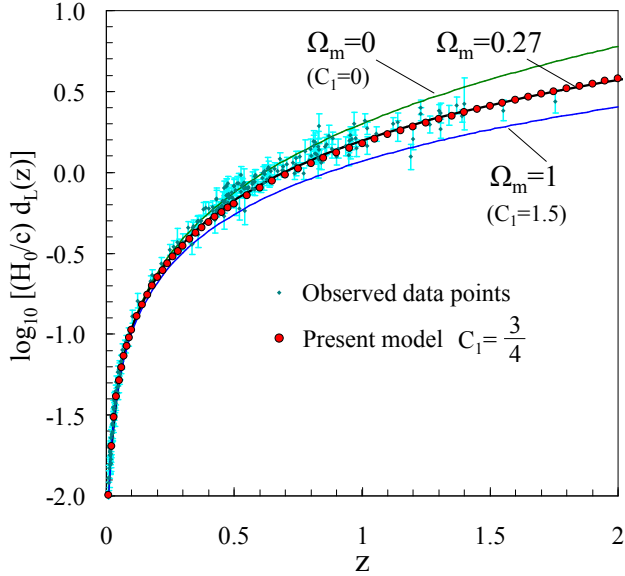}}
\end{center}
\end{minipage}
\caption{ (Color online). Dependence of luminosity distance $d_L$ on redshift $z$. 
The closed diamonds with error bars are supernova data points taken from Refs.\ \cite{Riess2007,SN1}. 
The closed circles represent the present simple model.  
The parameters are $w=0$, $\alpha_1=\beta_1=\gamma =1/2$, and $\alpha_2=\beta_2 = 0$, i.e., $C_{1}= 3/4$. 
For clarity, the present model is plotted as closed circles. 
The three solid lines represent the $\Lambda$CDM model for $(\Omega_{m}, \Omega_{\Lambda}) = (1, 0), (0.27, 0.73)$, and $(0, 1)$.
The solid lines for $\Omega_{m} = 1$ and $\Omega_{m}=0$ are equivalent to lines for $C_{1}=1.5$ and $C_{1}=0$.
Note that ($\Omega_{m}$, $\Omega_{\Lambda}$) is replaced by $\Omega_{m}$.
For supernova data points, $c$ and $H_{0}$ are set to be $3.0 \times 10^{5}$ [km/s] and $70$ [km/s/Mpc] \cite{WMAP2011}, respectively. }
\label{Fig-dL-z}
\end{figure}

Figure \ref{Fig-dL-z} shows the luminosity distance $d_{L}$ for the present simple model, supernova data points, and several $\Lambda$CDM models.  
We find that the simple model agrees well with supernova data points and the fine-tuned standard $\Lambda$CDM model, i.e., $(\Omega_{m}, \Omega_{\Lambda}) = (0.27, 0.73)$. 
It is successfully demonstrated that the simple model can mimic the present accelerating universe without adding the cosmological constant and dark energy. 
In entropic cosmology, at least, it is possible to discuss why $\gamma$ has such a value, unlike for the $\Lambda$CDM model. 
For example, $\gamma$ may be estimated from a derivation of surface terms or the Hawking temperature description \cite{Easson1}.
Note that Easson \textit{et al.} have suggested an alternate acceleration equation for a better fitting, where the entropic-force term is $\frac{3}{2\pi} H^{2} + \frac{3}{4\pi} \dot{H}$ \cite{Easson1}.  
In fact, we have confirmed that the present simple model is better than the alternate acceleration equation, 
i.e., the present model agrees well with the standard $\Lambda$CDM model, compared with the alternate acceleration equation suggested in Ref.\ \cite{Easson1}.

Next, we examine the entropy on the Hubble horizon, using the present simple model and $\Lambda$CDM models. 
The entropy $S$ for the simple model is calculated from Eq.\ (\ref{eq:SH-t}) with $C_{1}= 3/4$ for $\gamma = 1/2$, 
while $S$ for the $\Lambda$CDM model is calculated from $S=K/H^2$ as shown in Eq.\ (\ref{eq:SH}).
Since the entropies for $\Omega_{m}= 1$ and $0$ are equivalent to the entropies for $C_{1}=1.5$ and $0$, 
we can calculate them analytically.
On the other hand, we numerically compute the entropy for the fine-tuned standard $\Lambda$CDM model, i.e., $(\Omega_{m}, \Omega_{\Lambda}) = (0.27, 0.73)$. 
For the standard $\Lambda$CDM model, we employ the following equation \cite{Ryden1}: 
\begin{equation}
 H_{0} t = \int_{0}^{a} { \frac{  da }{   \sqrt {   \frac{ \Omega_{r} }{a^{2}}  + \frac{ \Omega_{m} }{a} + \Omega _{\Lambda} a^2  + ( 1- \Omega_{total} ) }   }    }    .
\label{eq:a_CDM}
\end{equation}
We first integrate Eq.\ (\ref{eq:a_CDM}) numerically, to obtain time evolution of the scale factor $a$.
The Hubble parameter $H$ is numerically calculated from $H=\dot{a}/a$.
Therefore, we can obtain the entropy $S=K/H^2$ for the standard $\Lambda$CDM model.
In the present study, we assume a spatially flat universe, i.e., $\Omega_{total}=1$, and neglect the influence of radiation, i.e., $\Omega_{r}=0$. 

\begin{figure} [t]  
\begin{minipage}{0.495\textwidth}
\begin{center}
\scalebox{0.32}{\includegraphics{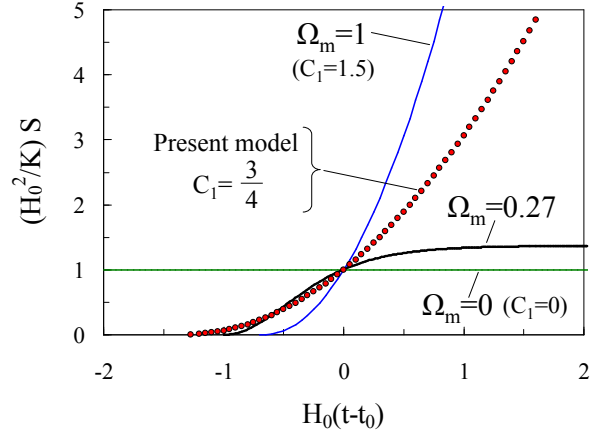}}
\end{center}
\end{minipage}
\caption{ (Color online). Time evolution of entropy $S$ on Hubble horizon.
The vertical and horizontal axes are normalized as $ (H_{0}^2 /K ) S $ and $H_{0} (t -t_{0})$, respectively. 
For details, see the caption of Fig.\ \ref{Fig-dL-z}.   }
\label{Fig-S-t_CDM}
\end{figure}

Now, we observe the time evolution of the entropy $S$ on the Hubble horizon.
As shown in Fig.\ \ref{Fig-S-t_CDM}, the entropies for both the simple model and the fine-tuned standard $\Lambda$CDM model increase with time until the present time, $H(t-t_{0}) \leq 0$.
In this sense, the simple model is similar to the fine-tuned standard $\Lambda$CDM model, i.e., $(\Omega_{m}, \Omega_{\Lambda}) = (0.27, 0.73)$. 
However, we can observe the difference between them clearly.
For example, the increase of the entropy for the simple model is likely uniform, even after the present time, i.e., $H(t-t_{0}) > 0$. 
In contrast, the increase of the entropy for the standard $\Lambda$CDM model tends to become gradually slower, especially after the present time, as if the present time were a special time.
This is because the cosmological constant $\Lambda$ is very dominant in the standard $\Lambda$CDM model.
To examine the entropy more closely, we focus on the rate of the change of the entropy. 
As shown in Fig.\ \ref{Fig-S-t_CDM}, $d^{2}S / dt^{2}$ for the simple model is always positive, 
while $d^{2}S / dt^{2}$ for the standard $\Lambda$CDM model is negative except for the early stage.
From Eq.\ (\ref{eq:SH-t}), we can confirm a positive $d^{2}S / dt^{2}$ for the present simple model,
because $d^{2}S / dt^{2} = 2 K C_{1}^{2} >0$ for $C_{1} \neq 0$, where $K$ is a positive constant given by Eq.\ (\ref{eq:K-def}).

As discussed above, the increase in entropy for the present simple model is uniform, while that for the fine-tuned standard $\Lambda$CDM model becomes gradually slower, especially after the present time.  
In other words, the standard $\Lambda$CDM model implies that the present time is a special time.
In fact, the entropy of the accelerated expanding universe has been extensively discussed from various viewpoints \cite{Easther1,Barrow3,Davies11,Davis01,Davis00,Jaeckel01,Gong00,Gong01,Setare1,Cline1}.
Many of the earlier works suggest that the increase of the entropy tends to become gradually slower, especially after the present time, because of the standard $\Lambda$CDM model. 
However, the simple model considered here predicts that the present time is not a special time, unlike for the prediction of the standard $\Lambda$CDM model.
[We have confirmed that the scale factor $a$ for the simple model increases uniformly even after the present time, 
while $a$ for the fine-tuned standard $\Lambda$CDM model increases rapidly after the present time. 
(The figure is not shown.)
This result is consistent with the result for the entropy shown in Fig.\ \ref{Fig-S-t_CDM}. 
This is because $S$ is calculated from $K/H^{2}$ [$=K/(\dot{a}/a)^{2}$], as shown in Eq.\ (\ref{eq:SH}).]

Finally, we  examine the deceleration parameter $q_{0}$. 
We can calculate $q_{0}$ for the present simple model as $ q_{0} = -0.25$, by substituting $C_{1}=3/4$ into $q_{0} = C_{1} -1 $ shown in Eq.\ (\ref{eq:q0C1}). 
On the other hand, $q_{0}$ for the standard $\Lambda$CDM model can be calculated as $q_{0} \thickapprox  -0.6$, from $q_{0} = (\Omega_{m} -2 \Omega_{\Lambda} + 2 \Omega_{r} ) /2 $ \cite{Weinberg1}, 
where $(\Omega_{m}, \Omega_{\Lambda}, \Omega_{r})$ is set to be  $(0.27, 0.73, 0)$. 
Therefore, at the present time, the acceleration for the simple model is slower than that for the standard $\Lambda$CDM model.
This is because the density parameter $\Omega_{\Lambda}$ for the cosmological constant is dominant in the standard $\Lambda$CDM model. 
In contrast, in the simple model, we assume the matter-dominated universe in entropic cosmology, without adding the cosmological constant and dark energy.

\section{Conclusions}

We have examined non-adiabatic expansion of the late universe and discussed the evolution of the entropy on the Hubble horizon, to study entropic cosmology from a thermodynamics viewpoint.
For this purpose, we have employed the two modified Friedmann equations, i.e., the modified Friedmann equation and the modified acceleration equation.
First of all, based on the two equations for entropic cosmology, we have examined the properties of the single-fluid dominated universe, neglecting high-order terms for quantum corrections. 
Consequently, we can systematically summarize the properties of the late universe, through a parameter $C_{1}$ related to entropic-force terms.
It is found that, at late times, the entropy on the Hubble horizon increases slowly with decreasing $C_{1}$ (or as the influence of entropic-force terms increases), 
while the expansion of the universe accelerates.

We have also derived the continuity equation from the first law of thermodynamics, assuming non-adiabatic expansion of the universe.
Using the obtained continuity equation, we have formulated the generalized Friedmann and acceleration equations,  
and have proposed a simple model as a possible model.
Through the luminosity distance, it is successfully shown that the simple model can explain the present accelerating universe and agrees well with both the supernova data and the fine-tuned standard $\Lambda$CDM model.
On the other hand, the increase of the entropy for the simple model is uniform,
although the increase of the entropy for the standard $\Lambda$CDM model is gradually slow especially after the present time. 
In other words, the simple model implies that the present time is not a special time, unlike for the prediction of the standard $\Lambda$CDM model.
We find that the present simple model predicts another future which is different from the standard $\Lambda$CDM model.

The present study has revealed the fundamental properties of the non-adiabatic expanding universe in entropic cosmology.
As one of several possible scenarios, the generalized formulation and the simple model considered here will help in understanding the accelerating expanding universe.
Of course, it is difficult to determine $\gamma$ related to the temperature on the Hubble horizon. 
However, at least in principle, it is possible to discuss the accelerating universe quantitatively, by means of the present entropic cosmology.  
Further discussions and observation data will be required to examine the present and future of the universe.

The modified continuity equation examined here has the so-called non-zero term on the right-hand side. 
Therefore, through the present paper, we call this the non-adiabatic process.
If we employ an effective description similar to bulk viscous cosmology, 
the non-zero term on the right-hand side can be cancelled in appearance.
Alternatively, the non-zero term may be interpreted as the interchange of energy between the bulk (the universe) and the boundary (the horizon of the universe). 
Accordingly, it will be necessary to study entropic cosmology from various viewpoints.

\appendix

\section{Bulk viscous cosmology}
\label{Appendix_Bulk}
In the present study, we consider a homogeneous and isotropic universe, 
and assume an entropy on the horizon of the universe for entropic cosmology.
However, usually, an only bulk viscosity can generate an entropy in the homogeneous and isotropic universe.
Such a cosmological model is called as `bulk viscous cosmology' and has been extensively investigated 
\cite{Davies3,Weinberg0,Murphy1,Barrow11,Barrow12,Zimdahl1,Arbab1,Brevik1,Brevik2,Nojiri1,Meng1,Fabris1,Colistete1,Barrow21,Meng2,Avelino1,Hipo1,Avelino2,Piattella1,Meng3,Pujolas1}.
In this appendix, we discuss similarities and differences between bulk viscous cosmology and entropic cosmology.

In bulk viscous cosmology, a bulk viscosity of cosmological fluids is assumed, and an effective pressure $p^{\prime}$ is given by 
\begin{equation}
 p^{\prime} = p - 3H \eta , 
\label{eq_Bulk_p}
\end{equation} 
where $\eta$ is the bulk viscosity \cite{Barrow12,Barrow21}. 
The continuity equation is 
\begin{equation}
 \dot{\rho} + 3 \frac{\dot{a}}{a}   \left (  \rho + \frac{p^{\prime}}{c^2}  \right ) = 0 . 
\label{eq_Bulk_Con0}
\end{equation}
Substituting Eq.\ (\ref{eq_Bulk_p}) into Eq.\ (\ref{eq_Bulk_Con0}) and arranging, we have  
\begin{equation}
 \dot{\rho} + 3 \frac{\dot{a}}{a} \left (  \rho + \frac{p}{c^2}  \right ) = \frac{ 9 H^2 \eta}{c^2}  .
\label{eq_Bulk_Con1}
\end{equation}
Equation\ (\ref{eq_Bulk_Con1}) is similar to Eq.\ (\ref{eq:fluid0}), because of a non-zero right-hand side.
The right-hand side of Eq.\ (\ref{eq_Bulk_Con1}) is related to a classical entropy generated by bulk viscous stresses. 
On the other hand, in entropic cosmology, 
we assume an entropy (and a temperature) on the horizon of the universe, instead of the classical entropy. 
Accordingly, the right-hand side of Eq.\ (\ref{eq:fluid0}) is related to the entropy on the horizon, unlike in bulk viscous cosmology. 
In fact, Davies \cite{Davies3} and Barrow \cite{Barrow12} have suggested a total entropy, 
i.e., the sum of the entropy on the horizon and the classical entropy generated by the bulk viscous stresses, to discuss the generalized second law of event horizon thermodynamics.
That is, in the present entropic cosmology, we focus on the entropy on the horizon, neglecting the classical entropy discussed above.  

We now examine the Friedmann equation.  
In bulk viscous cosmology, the Friedmann equation \cite{Barrow12} is given by 
\begin{equation}
 \left (  \frac{\dot{a}}{a} \right)^2 = \frac{8\pi G}{3} \rho   .         
\label{eq_Bulk_FRW1}
\end{equation}
Equation\ (\ref{eq_Bulk_FRW1}) does not include an additional term such as a cosmological constant.
On the other hand, in entropic cosmology, the Friedmann and acceleration equations have additional driving terms, 
which are derived from the usually neglected surface terms on the horizon. 
However, an additional term appears in the acceleration equation for bulk viscous cosmology. 
For example, the acceleration equation can be arranged as 
\begin{align}
\frac{\ddot{a}}{a}  
&=  - \frac{4\pi G}{3}  \left (  \rho + 3 \frac{p^{\prime}}{c^2}  \right )         \notag \\
&=  - \frac{4\pi G}{3}  \left (  \rho + 3 \frac{p}{c^2}  \right ) + \frac{12\pi G H \eta}{c^2} , 
\label{eq_Bulk_Accel}
\end{align}
where the last term, $12\pi G H \eta / c^2 $, corresponds to the additional driving term.
The additional term can explain the accelerated expansion of the universe.

\section{Dimensionless constants and an effective description for entropic cosmology}
\label{Appendix_F}
Entropic-force terms of the modified Friedmann and acceleration equations include four dimensionless constants $\alpha_{1}$, $\alpha_{2}$, $\beta_{1}$, and $\beta_{2}$.
In this appendix, we determine the dimensionless constants and examine an effective description for entropic cosmology.
To this end, we first derive the modified continuity equation from the modified Friedmann and acceleration equations, although we have already derived the modified continuity equation from the first law of thermodynamics.
This is because we can determine most of the dimensionless constants using the two continuity equations.

The modified Friedmann and acceleration equations, i.e., Eqs.\ (\ref{eq:mFRW01(H4=0)}) and (\ref{eq:mFRW02(H4=0)}), can be written as 
\begin{align}
\left(  \frac{ \dot{a}(t) }{ a(t) } \right)^2   
&=  \frac{ 8\pi G }{ 3 } \rho(t)  + \alpha_{1} H(t)^2  + \alpha_{2} \dot{H}(t)     \notag \\
&=  \frac{ 8\pi G }{ 3 } \rho(t)  + f(t)   ,
\label{eq:A_mFRW01}
\end{align}
and
\begin{align}
\frac{ \ddot{a}(t) }{ a(t) } 
&=  -  \frac{ 4\pi G }{ 3 } ( 1 +  3w) \rho(t)   + \beta_{1} H(t)^2  + \beta_{2} \dot{H}(t)      \notag \\
&=  -  \frac{ 4\pi G }{ 3 } ( 1 +  3w) \rho(t)   + g(t)   , 
\label{eq:A_mFRW02}
\end{align}
where $f(t)$ and $g(t)$ are given by 
\begin{equation}
f(t) = \alpha_{1} H(t)^2  + \alpha_{2} \dot{H}(t)   ,
\label{eq:A_f}
\end{equation}
and
\begin{equation}
g(t) = \beta_{1} H(t)^2  + \beta_{2} \dot{H}(t)      .
\label{eq:A_g}
\end{equation}
The four coefficients $\alpha_1$, $\alpha_2$, $\beta_1$, and $\beta_2$ are dimensionless constants. 
Equation\ (\ref{eq:A_mFRW01}) is the same formulation as Eq.\ (\ref{eq:mFRW01(f)}).
Therefore, as shown in Eq.\ (\ref{eq:dda_a}), arranging Eq.\ (\ref{eq:A_mFRW01}) gives 
\begin{equation}
\frac{ \ddot{a} }{ a } =  \frac{ 4\pi G }{ 3 } ( \dot{\rho} \frac{a}{\dot{a}} + 2 \rho ) + \frac{1}{2} \dot{f} \frac{ a }{ \dot{a} }  +  f    .
\label{eq:A_dda_a}
\end{equation}
Substituting Eq.\ (\ref{eq:A_dda_a}) into Eq.\ (\ref{eq:A_mFRW02}), and arranging this using $H = \dot{a} /a$, we have
\begin{equation}
       \dot{\rho} + 3  \frac{\dot{a}}{a}  ( 1   +  w )  \rho   
          =  \frac{3}{4 \pi G} H \left(  - f -  \frac{\dot{f} }{2 H }  +  g      \right )     . 
\label{eq:A_drho}
\end{equation}
When $f$ and $g$ are general functions, Eq.\ (\ref{eq:A_drho}) represents the generalized continuity equation.
However, in this appendix, $f$ and $g$ are given by Eqs.\ (\ref{eq:A_f}) and (\ref{eq:A_g}), respectively. 
Differentiating Eq.\ (\ref{eq:A_f}) with respect to $t$ gives 
\begin{equation}
\dot{f} = 2 \alpha_{1} H \dot{H}   + \alpha_{2} \ddot{H}   .
\label{eq:A_df}
\end{equation}
Accordingly, substituting Eqs.\ (\ref{eq:A_f}), (\ref{eq:A_g}), and (\ref{eq:A_df}) into Eq.\ (\ref{eq:A_drho}), we obtain 
\begin{align}
       & \dot{\rho} + 3  \frac{\dot{a}}{a}  ( 1   +  w )  \rho                                                        \notag \\
       &=  \frac{3}{4 \pi G}   \left [      (  -\alpha_{1} - \alpha_{2} + \beta_{2}) H \dot{H}        
                 +  (  -\alpha_{1} + \beta_{1}                 )  H^{3}   -   \frac{ \alpha_{2} }{2} \ddot{H}   \right ]    . 
\label{eq:A_fluid}
\end{align}
Equation\ (\ref{eq:A_fluid}) is the modified continuity equation derived from the modified Friedmann and acceleration equations.

We now determine the dimensionless constants as many as possible. 
For this purpose, we compare Eq.\ (\ref{eq:A_fluid}) with the modified continuity equation derived from the first law of thermodynamics,  i.e., Eq.\ (\ref{eq:fluid0r3}): 
\begin{equation}
       \dot{\rho} + 3  \frac{\dot{a}}{a}  ( 1 + w ) \rho  
          = - \gamma \left( \frac{3}{4 \pi G} H  \dot{H} \right )   . 
\label{eq:A_fluid_thermo}
\end{equation}
The two modified continuity equations, i.e., Eqs.\ (\ref{eq:A_fluid}) and (\ref{eq:A_fluid_thermo}), must be consistent with each other.
Therefore, three dimensionless constants can be determined when $H \dot{H}$, $H^{3}$, and $\ddot{H}$ are not $0$. 
The three constants in Eqs.\ (\ref{eq:A_mFRW01}) and (\ref{eq:A_mFRW02}) are given by
\begin{equation}
\alpha_{2} = 0   ,
\label{eq:A_const1}
\end{equation}
\begin{equation}
\beta_{1} = \alpha_{1}    , 
\label{eq:A_const2}
\end{equation}  
\begin{equation}
\beta_{2} = \alpha_{1} - \gamma . 
\label{eq:A_const3}
\end{equation}
Note that $\alpha_{1}$ and $\gamma$ should be determined from a different viewpoint, as mentioned previously.
Consequently, the modified self-consistent equations (i.e., the modified Friedmann, acceleration, and continuity equations) are summarized as  
\begin{equation}
 \left(  \frac{ \dot{a} }{ a } \right)^2    =  \frac{ 8\pi G }{ 3 } \rho  + \alpha_{1} H^2     ,  
\label{eq:A_mFRW01_4}
\end{equation}
\begin{equation}
\frac{ \ddot{a} }{ a }  =  - \frac{ 4\pi G }{ 3 } ( 1 + 3w)\rho   + \alpha_{1} H^2 +  (\alpha_{1} -\gamma) \dot{H} , 
\label{eq:A_accel4}
\end{equation}
\begin{equation}
       \dot{\rho} + 3  \frac{\dot{a}}{a}  ( 1   +  w )  \rho   
          = - \gamma \left( \frac{3}{4 \pi G} H  \dot{H} \right )     .
\label{eq:A_fluid4}
\end{equation}
Entropic-force terms of the modified Friedmann equation are only $H^{2}$-terms, as shown in Eq.\ (\ref{eq:A_mFRW01_4}).   

In the present study, we have selected $\alpha_{1} = \beta_{1} = \gamma $ and $\alpha_{2} = \beta_{2} = 0 $, to set up a simple model.
We can confirm that the selection is consistent with Eqs.\ (\ref{eq:A_const1}), (\ref{eq:A_const2}), and (\ref{eq:A_const3}).
Of course, the simple model is consistent with Eqs.\ (\ref{eq:A_mFRW01_4}), (\ref{eq:A_accel4}), and (\ref{eq:A_fluid4}), 
since $\gamma$ is selected as $\alpha_{1}$.
If Eqs.\ (\ref{eq:A_mFRW01}) and (\ref{eq:A_mFRW02}) are used for the modified Friedmann and acceleration equations, 
we can propose the above self-consistent equations to examine a non-adiabatic expansion of the late universe in entropic cosmology.

Finally, we examine an effective description for entropic cosmology discussed in Sec. \ref{Modified continuity equation}.
This is because it is possible to obtain an effective continuity (conservation) equation, 
when we employ an effective pressure similar to bulk viscous cosmology.
In this study, the effective pressure is given by  
\begin{equation}
   p^{\prime} = p + \frac{ \gamma c^2   }{4 \pi G} \dot{H}  , 
\label{eqA:pprime}
\end{equation}
and the equation of state parameter $w^{\prime}$ for the effective description is 
\begin{equation}
   w^{\prime}  =  \frac{ p^{\prime} }{ \rho c^{2} }              , 
\label{eqA:wprime}
\end{equation}
where $w^{\prime}$ is different from $w = p /( \rho c^{2} )$ of Eq.\ (\ref{eq:w}). 
We can arrange Eqs.\ (\ref{eq:A_accel4}) and (\ref{eq:A_fluid4}), using Eqs.\ (\ref{eqA:pprime}) and (\ref{eqA:wprime}).
As a result, the self-consistent equations based on the effective description are summarized as 
\begin{equation}
   \left(  \frac{ \dot{a} }{ a } \right)^2    =  \frac{ 8\pi G }{ 3 } \rho  + \alpha_{1} H^2   ,  
\label{eqA:mFRW01(f)3e}
\end{equation}

\begin{equation}
    \frac{ \ddot{a} }{ a }  =  - \frac{ 4\pi G }{ 3 } ( 1 + 3w^{\prime})\rho    + \alpha_{1} H^2  + \alpha_{1} \dot{H}   , 
\label{eqA:accel3e}
\end{equation}
\begin{equation}
       \dot{\rho} + 3  \frac{\dot{a}}{a}  ( 1   +  w^{\prime} )  \rho    = 0     . 
\label{eqA:fluid0r3e}
\end{equation}
As shown in Eqs.\ (\ref{eqA:accel3e}) and (\ref{eqA:fluid0r3e}), the effective description helps to simplify the formulas of entropic cosmology.
In particular, as shown in Eq.\ (\ref{eqA:fluid0r3e}), the so-called non-zero term on the right-hand side of Eq.\ (\ref{eq:A_fluid4}) is cancelled in appearance. 
Therefore, Eq.\ (\ref{eqA:fluid0r3e}) may be suitable for discussing the continuity equation. 
However, through the present paper, we employ Eq.\ (\ref{eq:A_fluid4}), to make the non-zero term clear.
[It should be noted that not only Eq.\ (\ref{eq:A_fluid4}) but also Eq.\ (\ref{eqA:fluid0r3e}) is different from the continuity (conservation) equation discussed by Easson \textit{et al.} \cite{Easson1}. 
Similarly, our dimensionless constants determined in this study are expected to be different from their suggested constants.]

\end{document}